\documentclass[twocolumn,showpacs,preprintnumbers,amsmath,amssymb]{revtex4}

\usepackage{epsfig}
\usepackage{subfigure}
\usepackage{bm}
\usepackage{graphicx}
\usepackage{dcolumn}
\usepackage{amsmath}
\graphicspath{{pics/}}    

\usepackage[normalem]{ulem} 
\usepackage[dvips]{color} 

\newcommand{\be}{\begin{equation}}
\newcommand{\ee}{\end{equation}}
\newcommand{\bea}{\begin{eqnarray}}
\newcommand{\eea}{\end{eqnarray}}

\newcommand{\UNIT}[1]{\mbox{$\,{\rm #1}$}}

\newcommand{\fm}{\UNIT{fm}}

\def\bi{\begin{itemize}}
\def\ei{\end{itemize}}

\def\fm3{$\mathrm{fm}^3$}
\def\f3{\mathrm{fm}^3}

\def\be{\begin{equation}}
\def\ee{\end{equation}}

\def\B0bar{$\bar{B^0}$}

\begin{document}

\title{Centrality dependence of the direct photon yield and elliptic
flow \\ in heavy-ion collisions at $\sqrt{s_{NN}}$ = 200 GeV}

\author{O.~Linnyk}
\email{Olena.Linnyk@theo.physik.uni-giessen.de}
\affiliation{%
 Institute for Theoretical Physics, %
  Justus Liebig University of Giessen, %
  35392 Giessen, %
  Germany %
}

\author{W.~Cassing}
\affiliation{%
 Institute for Theoretical Physics, %
  Justus Liebig University of Giessen, %
  35392 Giessen, %
  Germany %
}

\author{E.~L.~Bratkovskaya}%
\affiliation{%
 Institute for Theoretical Physics, %
 Johann Wolfgang Goethe University, %
 60438 Frankfurt am Main, %
 Germany; %
Frankfurt Institute for Advanced Studies, %
 60438 Frankfurt am Main, %
 Germany; %
}

\date{\today}

\begin{abstract}
We calculate the centrality dependence of direct photons produced in
Au+Au collisions at $\sqrt{s_{NN}}=200$~GeV  and their transverse
momentum spectra
within the Parton-Hadron-String Dynamics (PHSD) transport approach.
As sources for 'direct' photons, we incorporate the interactions of
quarks and gluons as well as hadronic interactions
($\pi+\pi\to\rho+\gamma$, $\rho+\pi\to\pi+\gamma$, meson-meson
bremsstrahlung $m+m\to m+m+\gamma$, meson-baryon bremsstrahlung
$m+B\to m+B+\gamma$), the decays of $\phi$ and $a_1$
 mesons and the photons produced in the initial hard collisions ('pQCD').
 We find that the $p_T$ spectra of the 'thermal' photons
 (i.e. the direct photons after the pQCD contribution is subtracted) deviate from exponential
 distributions and, consequently, observe a strong dependence of the inverse
slope parameter $T_{eff}$ on the fitting range in $p_T$. On the
other hand, all the obtained 'effective temperatures' are well above
the critical temperature for the deconfinement phase transition even
for peripheral collisions, reflecting primarily a 'blue shift' due
to radial collective motion of hadrons. Our calculations suggest
that the channel decomposition of the observed spectrum changes with
centrality with an increasing (dominant) contribution of hadronic
sources for more peripheral reactions. Furthermore, the thermal
photon yield is found to scale roughly with the number of
participant nucleons as $N_{part}^\alpha$ with $\alpha \approx$ 1.5,
whereas the partonic contribution scales with an
exponent  $\alpha_p \approx1.75$. Additionally, we provide predictions for the
centrality dependence of the direct photon elliptic flow $v_2(p_T)$.
The direct photon $v_2$ is seen to be larger in peripheral
collisions compared to the most central ones since the
photons from the hot deconfined matter in the early stages of the
collision carry a much smaller elliptic flow than the final
hadrons.
\end{abstract}


\maketitle

\section{Introduction}

The 'direct photons' from relativistic heavy-ion collisions are a valuable
probe of the collision dynamics
at early times and provide information on the characteristics of the
initially created matter once the final state hadronic decay photons
are subtracted from the experimental
spectra~\cite{Shuryak:1977ut,Shuryak:1978ij,Feinberg:1970tg,Feinberg:1976ua,Bjorken:1975dk,Peitzmann:2001mz,Aurenche:2002yi}.
In the last years, the PHENIX
Collaboration~\cite{PHENIX1,PHENIXlast,Adare:2008ab,Tserruya:2012jb}
has measured the spectra of the photons produced in minimal bias
Au+Au collisions at $\sqrt{s_{NN}}=200$~GeV and found a strong
elliptic flow $v_2(p_T)$ of 'direct photons', which is comparable to
that of the produced pions. Since direct photons were expected to be
essentially produced in the initial hot medium before the collective
flow has developed, this observation was in contrast to the
theoretical expectations and predictions
\cite{Chatterjee:2005de,Liu:2009kq,Dion:2011vd,Dion:2011pp,vanHees:2011vb}.
Also more recent studies employing  event-by-event hydrodynamical
calculations~\cite{Chatterjee:2013naa,Shen:2013vja,Shen:2013cca}
severely have underestimated the elliptic flow of direct photons and
alternative sources of direct photons from the conformal anomaly
have been suggested~\cite{Basar:2012bp}. Furthermore, in order to
distinguish direct photons from the strong magnetic field of
spectator protons (due to the conformal anomaly) it has been
suggested to explore the centrality dependence of the direct photon
$v_2$ in correlation with the elliptic flow from
pions~\cite{Bzdak:2012fr}.

On the other hand, in Ref.~\cite{Linnyk:2013hta} we have proposed
that apart from the partonic production channels the direct photon
yield and primarily the strong $v_2$ might be due to hadronic
sources (such as meson-meson Bremsstrahlung or hadronic interactions
as $\pi+\pi\to\rho+\gamma$, $\rho+\pi\to\pi+\gamma$ etc.).  Indeed,
the interacting hadrons carry a large $v_2$ and contribute by
more than 50\% to the measured 'direct photons' according
to the PHSD calculations in Ref.~\cite{Linnyk:2013hta} (cf. also the
hydrodynamics calculations in Ref.~\cite{Dusling:2009ej}). For a
quantitative understanding of the direct photon production it is
important to verify the decomposition of the total photon yield
according to the production sources: the late hadron decays (the
cocktail), hadronic interactions beyond the cocktail (during the
collision phase) and the partonic interactions in the quark-gluon
plasma (QGP). Since previous transport studies have indicated that
the duration of the partonic phase substantially decreases with
increasing impact parameter (cf. Fig.~4 in
Ref.~\cite{PHSDasymmetries}) we will study here explicitly the
centrality dependence of the direct photon yield together with the
essential production channels and their impact on the photon $v_2$.

As in Ref.~\cite{Linnyk:2013hta} we will employ the
Parton-Hadron-String Dynamics (PHSD) transport approach to
investigate the photon production in Au+Au collisions at
$\sqrt{s_{NN}}=200$~GeV at various centralities thus extending our
previous investigations for the case of minimum bias collisions.
We recall that the PHSD approach has provided a consistent
description of the bulk properties of heavy-ion collisions --
rapidity spectra, transverse mass distributions, azimuthal
asymmetries of various particle species -- from low
Super-Proton-Synchrotron (SPS) to top
Relativistic-Heavy-Ion-Collider (RHIC) energies~\cite{PHSDqscaling}
and was successfully used also for the analysis of dilepton
production from hadronic and partonic sources at SPS, RHIC and
Large-Hadron-Collider (LHC) energies ~\cite{Linnyk2011_BOTH}. It is
therefore of interest to calculate also the photon production in
relativistic heavy-ion collisions from  hadronic and partonic
interactions within the PHSD transport approach, since its
microscopic and non-equilibrium evolution of the nucleus-nucleus
collision is independently controlled by a multitude of other
hadronic and electromagnetic observables in a wide energy
range~\cite{BrCa11,PHSDasymmetries,Konchakovski:2011qa,Linnyk2011_BOTH}.

\section{Photons within PHSD}
For the details on the PHSD approach we refer the reader to Refs.
\cite{BrCa11,CasBrat} and the implementation of the photon
production to Refs.~\cite{ElenaKiselev,Linnyk:2013hta} (and
references therein). Let us recall that the dynamical calculations
within the PHSD have reproduced the measured differential spectra of
dileptons produced in heavy-ion collisions at SPS and RHIC energies
(see Refs.~\cite{Linnyk2011_BOTH}). Furthermore,
the dilepton production rate from the QGP constituents - as incorporated in
the PHSD - agrees with the dilepton rate from the thermalized
QCD medium as calculated  by lattice QCD (lQCD). We note,
additionally, that the electric conductivity of the QGP from the
PHSD, which controls the photon emission rate in equilibrium, is
rather well in line with available lQCD results
\cite{Cassing:2013iz}.

\begin{figure}
\resizebox{0.47\textwidth}{!}{%
 \includegraphics{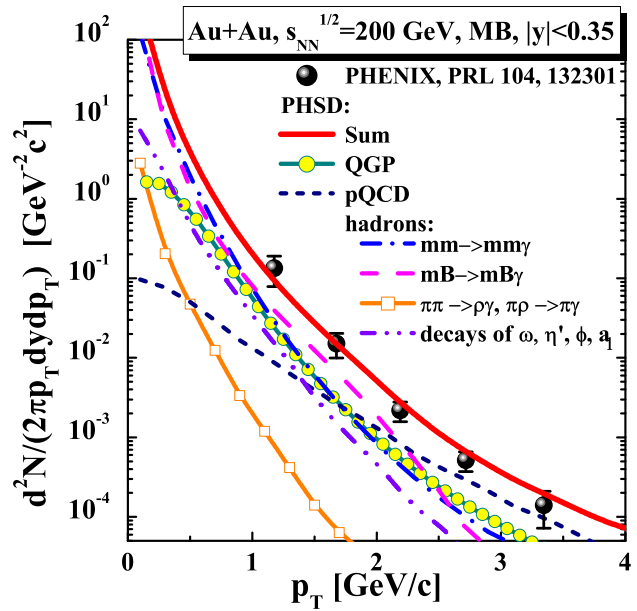}
} \caption{(Color online) Direct photons (sum of all photon
production channels except the $\pi$- and $\eta$-meson decays) from
the PHSD approach (red solid line) in comparison to the data of the
PHENIX Collaboration~\protect{\cite{PHENIXlast,Adare:2008ab} for
minimal bias collisions of Au+Au at $\sqrt{s_{NN}}=200$~GeV } (black
symbols). The various channels are described in the legend. }
\label{MB}
\end{figure}

As sources of photon production - on top of the general dynamical
evolution - we consider
hadronic~\cite{Turbide:2003si,ElenaKiselev,Kapusta:1991qp,Gale87} as
well as
partonic~\cite{Feinberg:1976ua,Shuryak:1978ij,Linnyk:2004mt,olena2010,Wong:1998pq}
interactions. In the present study we extend the calculations in
Ref.~\cite{Linnyk:2013hta} by adding an additional source of
photons, i.e. the bremsstrahlung in elastic meson+baryon collisions
($m+B\to m+B+\gamma$). In our previous study
(Ref.~\cite{Linnyk:2013hta}), we have considered the meson-meson
bremsstrahlung, because it had been proposed to be important already
in Refs.~\cite{Gale87,ElenaKiselev}. At the time we had not realized
the potential importance of the photon production in meson-baryon
collisions. However, we will see below that this process contributes
considerably.

The bremsstrahlung production of photons is calculated using the
soft photon approximation as in
Refs.~\cite{Gale87,PhysRevD.53.4822,ElenaKiselev,Linnyk:2013hta}.
The soft-photon approximation (SPA) relies on the assumption that
the radiation from internal lines is negligible and the strong
interaction vertex is on-shell. In this case the strong interaction
part and the electromagnetic part can be separated, so the
soft-photon cross section for the reaction $h_1 + h_2 \to h_1 + h_2
+ \gamma$ can be written as
\begin{eqnarray}
q_0\frac{d^3 \sigma^\gamma}{d^3 q} & = & \frac{\alpha}{4 \pi}
\frac{{\bar \sigma (s)}}{q_0^2},
            \label{brems} \\
{\bar \sigma(s)} & = & \frac{s - (M_1 + M_2)^2}{2 M_1^2} \sigma(s),
\nonumber
\end{eqnarray}
where $M_1$ is the mass of the charged accelerated particle, $M_2$
is the mass of the second particle; $q_0, q$ are the energy and
momentum of the photon. In (\ref{brems})  $\sigma(s)$ is the
on-shell cross section for the reaction $h_1+h_2\to h_1+h_2$, i.e.
the elastic scattering of the two hadrons.

Let us point out that the resulting yield of the bremsstrahlung
photons depends on the model assumptions such as the cross sections
for the meson-meson and meson-baryon elastic scatterings,
incoherence of the individual scatterings and the soft photon
approximation. The theoretical uncertainty of up to a factor of 2
due to the unmeasured elastic scattering cross sections has to be
kept in mind. The adequacy of the SPA assumption has been checked in
Ref.~\cite{PhysRevD.53.4822}. We recall here that the soft photon
approximation is no longer valid for high energies of the produced
photons or at high $\sqrt{s}$ of the meson+meson or meson+baryon
collisions~\cite{Eggers:1996bs}. Thus we have restricted our
kinematics by considering only meson+meson and baryon+meson
collisions with available energies $\sqrt{s}$ below 3~GeV. Our
conclusions on the centrality dependence of the direct photons are
not sensitive to the actual value of the cut-off within reasonable
variations.

The assumption of incoherent photon production in individual
hadron-hadron collisions is not applicable at very low transverse
momenta of the photons. The Landau-Pomeranchuk-Migdal (LPM) effect
is the suppression of bremsstrahlung photon emission due to the
multiple scattering of the production source (in this case meson or
baryon) during the time needed for the formation of the radiated
photon. In this case the bremsstrahlung amplitudes interfere
destructively. For the hadronic bremsstrahlung, the LPM effect in
the thermal medium has been calculated in
Ref.~\cite{Cleymans:1992kb}. The suppression depends on the length
of the formation zone of the photon $z(\gamma)$, which is defined by
the uncertainty principle and depends on the energy of the photon.
The suppression becomes significant for photon energies below a
certain value, for which $z(\gamma)$ becomes larger than the mean
free path of the hadron $\lambda$~\cite{Anthony:1997ed,Wang:1994fx}.
For the photon at mid-rapidity, $z(\gamma)=2 \omega/p_T^2=2/p_T$. On
the other hand, the mean free path of the hadrons $\lambda=1/(\sigma
n)$ is governed by the hadronic scattering cross section (typically
of the order of $\sigma=20$ mb) and the hadron density, which after
the hadronization does not exceed $n_{max}=0.5$~fm$^{-3}$.
Accordingly, the suppression due to the LMP effect is expected to be
important for these processes at $p_T<0.4$ GeV, where, however, no
data are available yet. At present, we do not include the LMP effect
on the bremsstrahlung photon production in our calculations due to
the lack of data at sufficiently low $p_T$.

\begin{figure}
\resizebox{0.47\textwidth}{!}{%
 \includegraphics{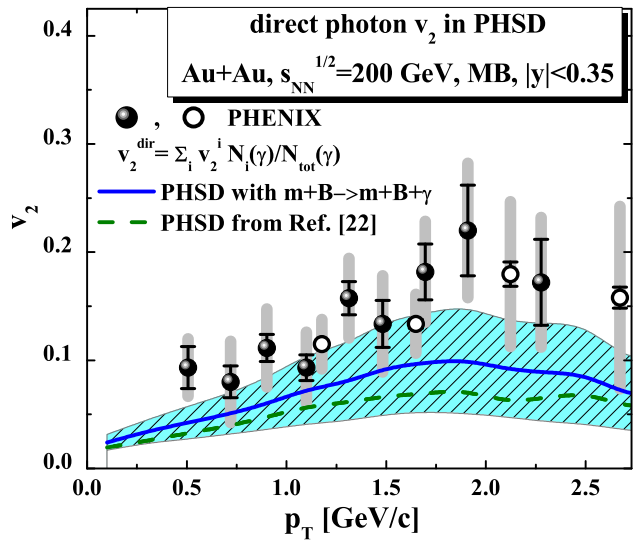}
} \caption{(Color online) Direct photon elliptic flow (contributions
from hadronic decays are subtracted) from the PHSD approach in
comparison to the data of the PHENIX
Collaboration~\protect{\cite{PHENIXlast,Adare:2008ab}} for minimal
bias collisions of Au+Au at $\sqrt{s_{NN}}=200$~GeV (black symbols).
The green dashed line shows the PHSD results from the
Ref.~\protect{\cite{Linnyk:2013hta}} taking into account the
following channels: $\pi+\rho\to\pi+\gamma$, $\pi+\pi\to
\rho+\gamma$; the photon bremsstrahlung in meson-meson collisions
$m+m\to m+m+\gamma$; photon production in the QGP in the processes
$q+{\bar q}  \to g+\gamma$, and $q({\bar q})+g \to q({\bar
q})+\gamma$ as well as the photon production in the initial hard
collisions ("pQCD"). The blue solid line give the results of the
present calculations taking into account additionally the
baryon-meson bremsstrahlung $m+B\to m+B+\gamma$.} \label{MBv2}
\end{figure}

\begin{figure*}
\includegraphics[width=0.95\textwidth]{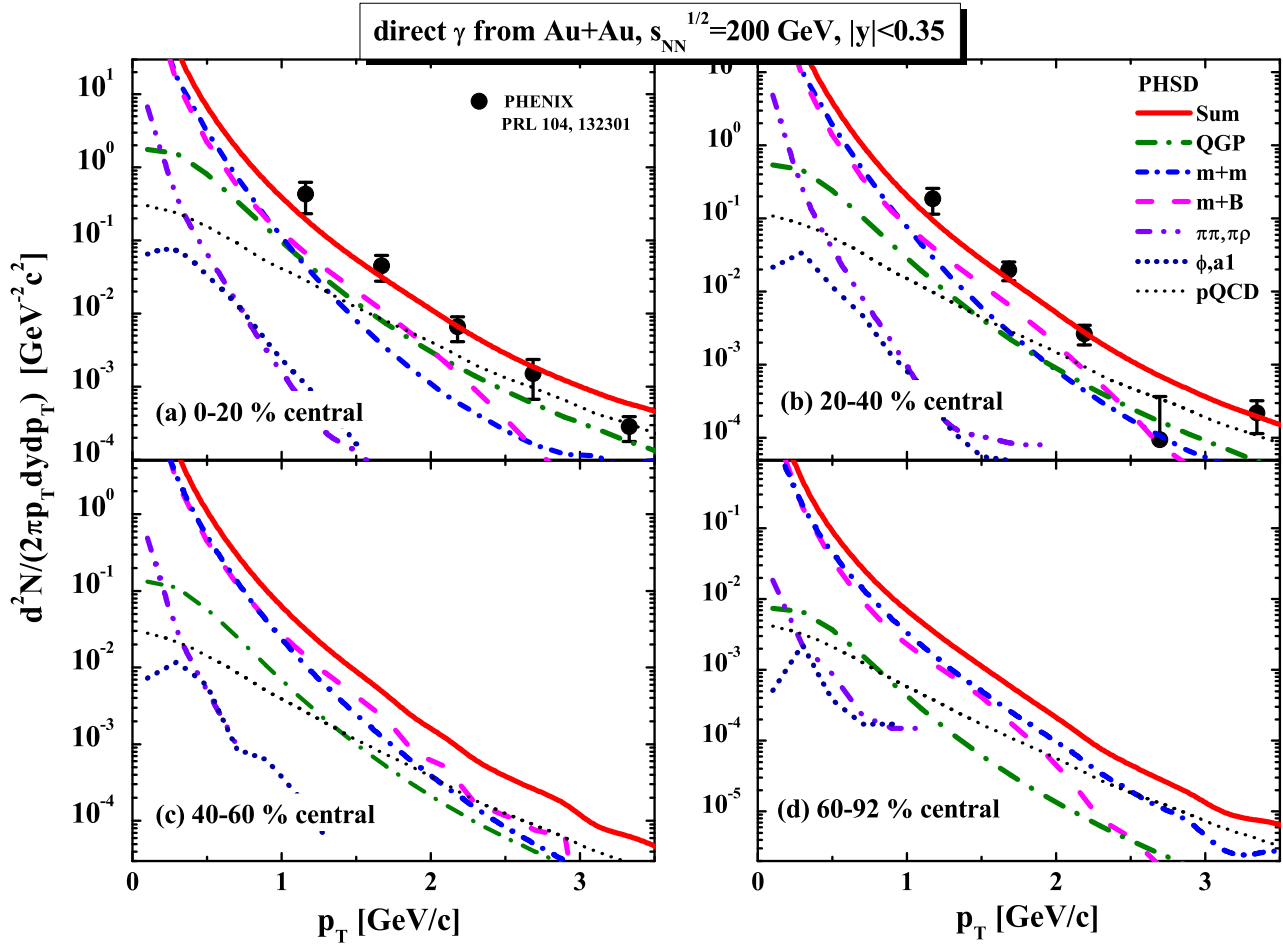}
\caption{(Color on-line) The channel decomposition of the direct
photon transverse momentum ($p_T$) spectra for Au+Au collisions at
$\sqrt{s_{NN}}=200$~GeV (full solid upper line) at mid-rapidity
$|y|< 0.35$ within the PHSD approach. The four panels present the
results at various collision centralities: (a) 0-20 \% central, (b)
20-40 \% central, (c) 40-60 \% central and (d) 60-92 \%. The channel
description is given in the legend. The data are
from~\protect{\cite{Adare:2008ab}}.} \label{4_spectra}
\end{figure*}

Since a new production mechanism has been added to the hadronic
production channels (the meson-baryon bremsstrahlung), we first
check whether this addition does not lead to an overestimation of
the data from the PHENIX
Collaboration~\cite{PHENIXlast,Adare:2008ab}  in minimal bias
$Au+Au$ collisions in Fig.~\ref{MB}. Since the decays of mesons as
'late' hadronic sources are less sensitive to the creation of the
hot and dense medium and to its properties, they are usually
subtracted experimentally from the total photon yield to access the
`direct' photon spectrum. In our calculations of the direct photon
spectrum in Fig.~\ref{MB} the following sources are taken into
account: the decays of $\omega$, $\eta$', $\phi$ and $a_1$ mesons;
the reactions $\pi+\rho\to\pi+\gamma$, $\pi+\pi\to \rho+\gamma$; the
photon bremsstrahlung in meson-meson and meson-baryon collisions
$m+m\to m+m+\gamma$, $m+B\to m+B+\gamma$; photon production in the
QGP in the processes $q+{\bar q}  \to g+\gamma$, and $q({\bar q})+g
\to q({\bar q})+\gamma$ as well as the photon production in the
initial hard collisions ("pQCD"), which is given by the hard photon
yield in p+p collisions scaled with the number of binary collisions
$N_{coll}$. We find that our PHSD calculations are in a reasonable
agreement with the PHENIX data~\cite{PHENIXlast,Adare:2008ab} and
show a clear dominance of the hadronic production channels over the
partonic channels for transverse momenta below about 0.7 GeV/c. In
particular, the bremsstrahlung contributions are responsible for the
'banana shape' spectrum and the strong increase for low $p_T$. On
the other hand, this increase should be softened to some degree by
the LPM effect. Accordingly, especially experimental data well below
1 GeV/c in $p_T$ will be helpful in disentangling the various
sources.

In Fig.~\ref{MBv2}, we show explicitly the elliptic flow $v_2$ of
direct photons in minimum bias collisions in comparison to the data
and the previous centrality integrated results also for  (the green
dashed line) from the Ref.~\cite{Linnyk:2013hta}.  Note that the
photons from the decays of $\omega$ and  $\eta$' mesons have been
subtracted from the $v_2$ data by experimental methods. We
calculated the direct photon $v_2$ as a sum of $v_2(i)$ of each
individual contributed channel, weighted with the channel's
contribution to the $p_T$ spectrum. The considered channels are:
$\pi+\rho\to\pi+\gamma$, $\pi+\pi\to \rho+\gamma$; the photon
bremsstrahlung in meson-meson collisions $m+m\to m+m+\gamma$; photon
production in the QGP in the processes $q+{\bar q}  \to g+\gamma$,
and $q({\bar q})+g \to q({\bar q})+\gamma$ as well as the photon
production in the initial hard collisions ("pQCD"). The new
calculations include additionally the meson bremsstrahlung processes
$m+B\to m+B+\gamma$ and are shown by the blue solid line. The
agreement with experiment has slightly improved compared to
Ref.~\cite{Linnyk:2013hta}.

\begin{figure*}
\resizebox{0.95\textwidth}{!}{%
 \includegraphics{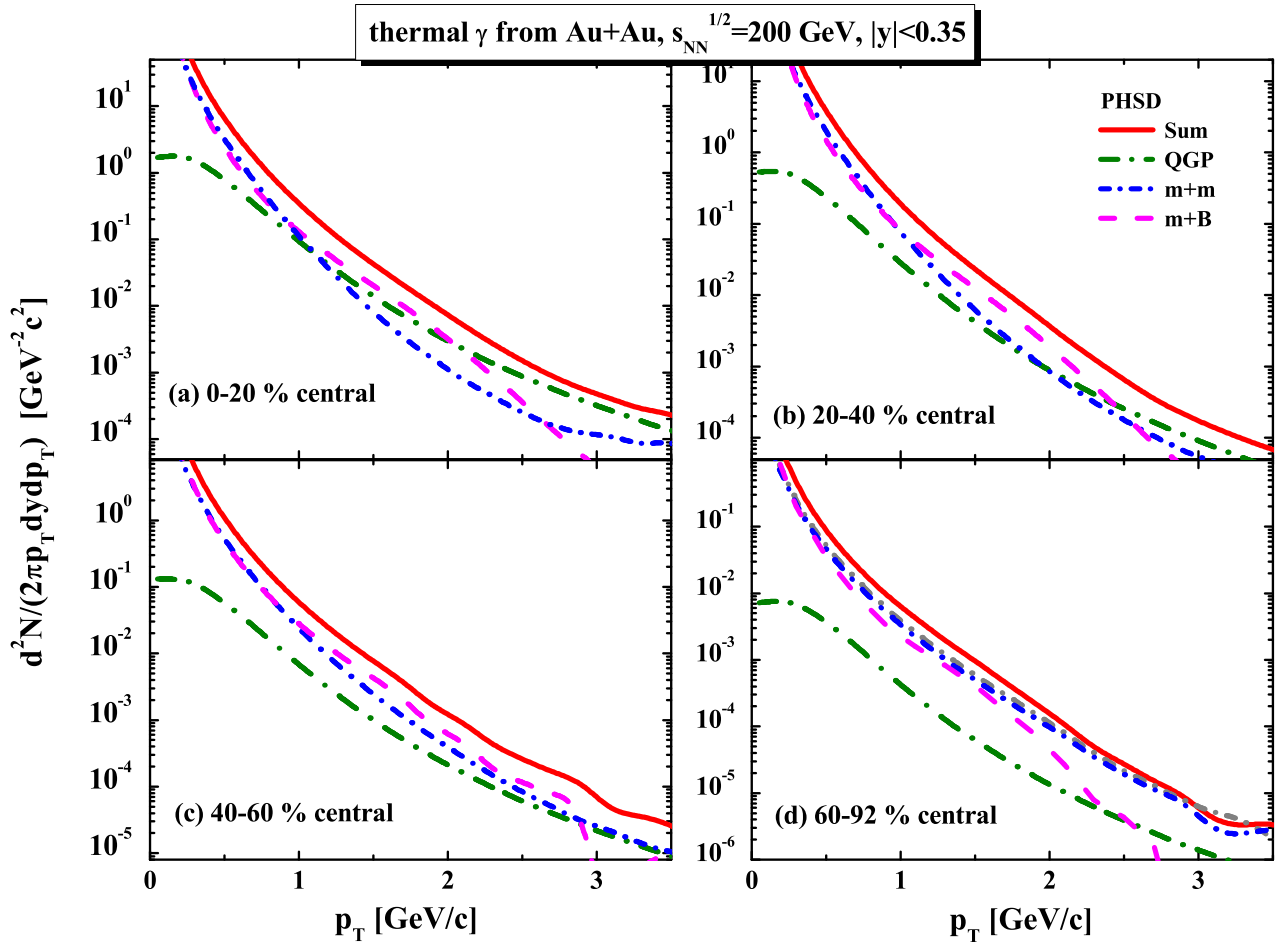}
} \caption{(Color online) The spectra of thermal photons in (a) 0-20
\% central, (b) 20-40 \% central, (c) 40-60 \% central and (d) 60-92
\% central Au+Au collisions at $\sqrt{s_{NN}}=200$~GeV within the
PHSD approach. In contrast to the plots in Fig.~\ref{4_spectra}, we
subtracted the contribution from the initial hard partonic
collisions (the ``pQCD" channel).} \label{1_plot}
\end{figure*}

\section{Results for different centralities}

The calculated results for the direct photon spectrum in Au+Au
collisions at $\sqrt{s_{NN}}=200$~GeV  are presented in
Fig.~\ref{4_spectra} for various centralities as functions of the
transverse momentum $p_T$ at mid-rapidity $|y|< 0.35$. The direct
photons are obtained experimentally from the total photon spectrum
by subtracting meson decay photons base on measured meson yields.
Therefore, in this case we disregard all hadron decays except the
$\phi, a_1$ photonic decays, which are subleading. The following
contributions are addressed as 'direct photons': the decays of
$\phi$ and $a_1$ mesons; the reactions $\pi+\rho\to\pi+\gamma$,
$\pi+\pi\to \rho+\gamma$; the photon bremsstrahlung in meson-meson
and meson-baryon collisions $m+m\to m+m+\gamma$, $m+B\to
m+B+\gamma$; photon production in the QGP in the processes $q+{\bar
q}  \to g+\gamma$, and $q({\bar q})+g \to q({\bar q})+\gamma$ as
well as the photon production in the initial hard collisions
("pQCD"). The direct pQCD contributions dominates above $p_T
\approx$ 2 GeV/c.

The spectra of ``thermal" photons are obtained from the direct
photon spectra (channels listed above) by additionally subtracting
the photons produced in the initial hard pQCD processes. The pQCD
photons are not expected to have thermal spectrum and practically
give no contribution to the direct photon elliptic flow. The thermal
photons in 0-20 \% central, 20-40 \% central, 40-60 \% central and
60-92 \% central Au+Au collisions at $\sqrt{s_{NN}}=200$~GeV within
the PHSD approach are displayed in Fig.~\ref{1_plot}. We only
specify the dominant channels in Fig.~\ref{1_plot}, i.e. the
contributions from m+m and m+B bremsstrahlung as well as the QGP
contribution which is seen to become low in more peripheral
collisions.

Though the spectrum presented in Fig.~\ref{1_plot} is obviously not
exponential in the full momentum range especially due to the
bremsstrahlungs channels, one may fit the spectra by exponentials in
a finite transverse momentum region and define in this way an
effective slope parameter or `effective temperature' as in the
experimental analysis~\cite{PHENIXlast,Adare:2008ab}. The slope of
the transverse momentum spectrum of produced 'thermal photons' is
expected to give a glance at the initial temperatures reached in the
collisions~\cite{Chatterjee:2005de,Liu:2009kq,Dion:2011vd,Dion:2011pp,Feinberg:1976ua,Shuryak:1978ij,Kapusta:1991qp,Alam:2000bu,
Steffen:2001pv,Srivastava:2000pv,Huovinen:2001wx,Turbide:2003si,d'Enterria:2005vz,Liu:2008eh,Turbide:2007mi,Shen:2013vja},
and was even used to deduce an 'average temperature' of the
QGP~\cite{PHENIXlast,Adare:2008ab}.  We will present here the
effective temperatures $T_{eff}$ as extracted from the calculated
transverse momentum spectra of thermal photons from
Fig.~\ref{1_plot}, addressing them as 'apparent inverse slope
parameters' $T_{eff}$ or energy scales for the photonic radiation.
 The extracted `effective temperatures' are shown in Table I at
 the different centralities for the
transverse momentum interval 0.6-2~GeV. The `temperature' defined in
this way depends on the fit range in transverse momentum and should
only serve as a characteristic energy scale as mentioned above.
Surprisingly, we find (within error bars) the same slope parameter
$T_{eff}$ which is significantly larger than the critical
temperature $T_c \approx$ 160 MeV for deconfinement. Since here the
dominant contributions should be related to binary bremsstrahlung
channels the high slope parameters predominantly reflect the
'blue-shift' of the photon spectra due to the collective flow of
hadrons (cf. Ref.~\cite{Shen:2013vja}) which (for PHSD) was shown in
Ref.~\cite{PHSDasymmetries} to be well in line with experimental
observation.
\begin{table}[h]
\begin{tabular}{|c|c|c|}
\hline
  \multicolumn{3}{|c|}{ \phantom{\Large I} The slope parameter $T_{eff}$ (in MeV) \phantom{\Large I}} \\
\hline \phantom{\Large I} Centrality \phantom{\Large I} &
\phantom{\Large I} $N_{part}$ \phantom{\Large I} & \phantom{\Large
I} $T_{eff}$
\\ \hline \ 0-20\% \ & \ 280 \ &
\phantom{\Large I} $265\pm20$ \
\\ \hline \ 20-40\% \ & \ 137 \ &
\phantom{\Large I} $260\pm20$ \
\\ \hline \ 40-60\% \ & \ 60 \ &
\phantom{\Large I} $250\pm20$ \
\\ \hline \ 60-92\% \ & \ 15 \ &
\phantom{\Large I} $260\pm20$ \
\\ \hline
\end{tabular}
\caption{The slope parameter $T_{eff}$ of the spectrum of thermal
photons (Fig.~\ref{1_plot}) produced in Au+Au collisions at
$\sqrt{s_{NN}}=200$~GeV at various centralities. The value $T_{eff}$
was obtained by approximating the spectrum by an exponential
function in the transverse momentum range $0.6<p_T<2$~GeV.}
\label{table}
\end{table}

\begin{figure}
\resizebox{0.47\textwidth}{!}{%
 \includegraphics{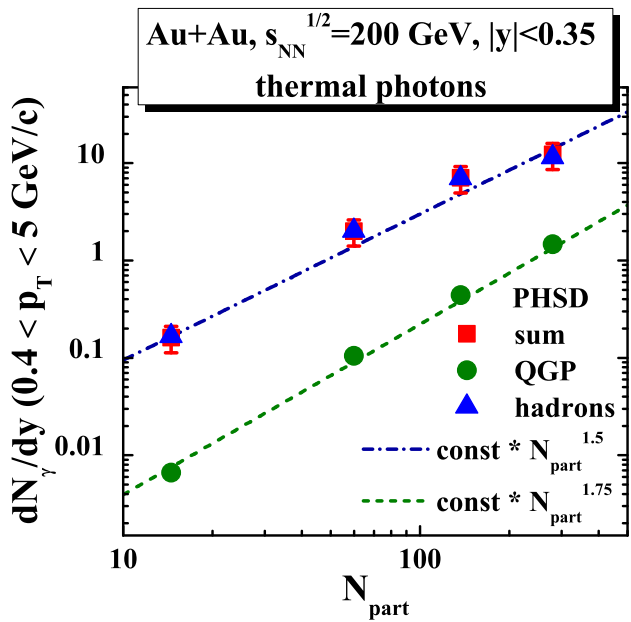}
} \caption{(Color online) Integrated spectra of thermal photons
(full squares) produced in Au+Au collisions at
$\sqrt{s_{NN}}=200$~GeV versus
 the number of participants
$N_{part}$. The scaling with $N_{part}$ from the QGP contribution
(full dots) and the bremsstrahlungs channels (full triangles) are
shown separately.} \label{np}
\end{figure}

Integrating the thermal photon spectra from Fig.~\ref{1_plot} over
the transverse momentum $p_T$ in the interval $0.4 \leq p_T \leq 5$,
we obtain the number of thermal photons (full squares) as a function
of centrality, which is plotted in Fig.~\ref{np} as a function of
the number of participants $N_{part}$  calculated in the Monte-Carlo
Glauber model described in Ref.~\cite{Miller:2007ri}. Since only
binary collision channels contribute to the production of thermal
photons in our approach, their yield rises faster than $N_{part}$ as
expected from qualitative considerations in
Refs.~\cite{Kajantie:1986dh,Srivastava:1999jz}. A power-law fit to
our results gives approximately a scaling $\sim N_{part}^\alpha$
with $\alpha \approx$ 1.5. In addition we display in Fig.~\ref{np}
the scaling with $N_{part}$ for the partonic (full dots) and
hadronic bremsstrahlung channels (full triangles) separately, which
give exponents of $\approx$ 1.75 and $\approx$ 1.5, respectively.

\begin{figure}
\resizebox{0.47\textwidth}{!}{%
 \includegraphics{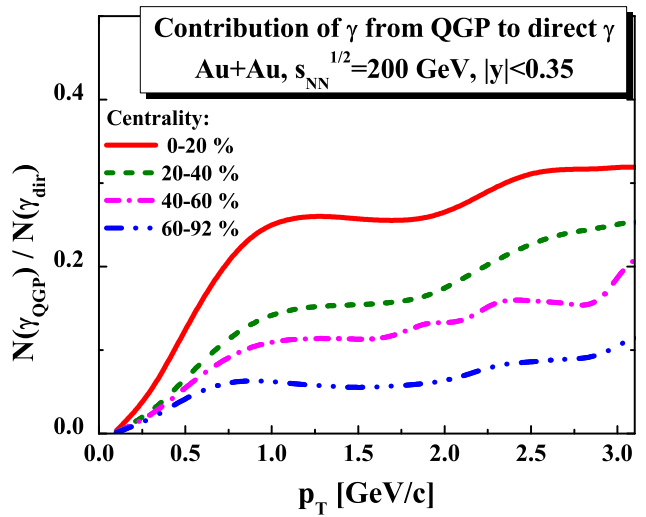}
} \caption{(Color online) The ratios of the number of photons
produced in the QGP to the number of all direct photons produced
through binary processes in different-centrality Au+Au collisions at
$\sqrt{s_{NN}}=200$~GeV versus the photon transverse momentum $p_T$.}
\label{ratio}
\end{figure}

As one can see in Figs.~\ref{4_spectra} and~\ref{1_plot}
qualitatively  the contribution of the photons from the QGP is
larger in central collisions while the hadronic sources contribute
more dominantly in peripheral collisions. We quantify the relative
contributions by plotting in Fig.~\ref{ratio} the ratio of the
number of photons produced in the QGP to the number of  all direct
photons (from the QGP, $m+m/B\to m+m/B+\gamma$,
$\pi+\pi/\rho\to\rho/\pi+\gamma$ and the pQCD photons). The
contribution of the QGP photons is seen to increase with transverse
momentum and reaches slightly more than 30\% for the most central
event bin. On the other hand, the ratio of QGP photons to the total
direct photons falls rapidly with decreasing centrality and is below
10\% in the most peripheral centrality bin. Accordingly, minimal
bias collisions are dominated by the hadronic channels that come
along with a large hadronic elliptic flow $v_2$.

\begin{figure}
\resizebox{0.47\textwidth}{!}{%
 \includegraphics{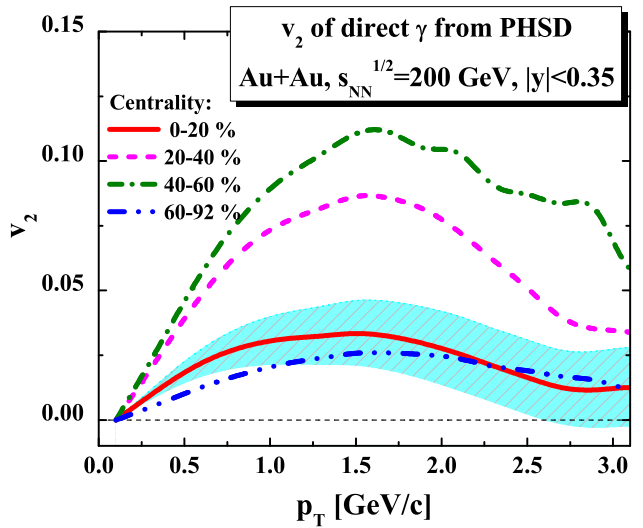}
} \caption{(Color online) The elliptic flow $v_2(p_T)$ of direct photons
produced through binary processes in  Au+Au collisions at
$\sqrt{s_{NN}}=200$~GeV for different centralities versus the photon transverse momentum $p_T$.
The hatched area (for the most central bin) stands for the statistical uncertainty in the photon $v_2$
from PHSD which in width is also characteristic for the other centralities. }
\label{v2}
\end{figure}

In Fig.~\ref{v2} we provide predictions for the centrality
dependence of the direct photon elliptic flow $v_2(p_T)$ within the
PHSD approach. The direct photon $v_2$ is seen to be larger in the
peripheral collisions compared to the most central ones. The
predicted centrality dependence of the direct photon flow results
from the interplay of two independent factors. Firstly, the channel
decomposition of the direct photon yield as presented by the ratios
in Fig.~\ref{ratio} changes: the admixture of photons from the
hadronic phase increases for more peripheral collisions. As has been
described in detail in Ref.~\cite{Linnyk:2013hta}, the PHSD approach
predicts a very small $v_2$ of photons produced in the initial hot
deconfined phase by partonic channels  of the order of 2\%. On the
other hand, the photons from the hadronic sources show strong
elliptic flow (up to 10\%), on the level of the $v_2$ of final
hadrons~\cite{Linnyk:2013hta}. Accordingly, since the channel
decomposition of the direct photons changes with centrality, the
elliptic flow of the direct photons increases with decreasing
centrality and becomes roughly comparable with the elliptic flow of
pions in peripheral collisions.

However, there is another (second) factor contributing to the
centrality dependence of the photon elliptic flow. Let us recall the
centrality dependence of the elliptic flow for charged particles,
e.g. Fig.7 of Ref.~\cite{PHSDasymmetries}. The $v_2$ rises almost
linearly with increasing impact parameter $b$ at small $b$, and
decreases again in the most peripheral collisions. The latter
decrease is a sign that the most peripheral collisions can be
understood as rather a superposition of elementary collisions, with
little collectivity. The elliptic flow in the most peripheral bin is
low in Fig.~~\ref{v2}, because the all the particles have little
flow at this high $b$. This effect is present in the PHSD model as
well as in the observation.

\section{Summary}

The spectra of direct and thermal photons - as produced in Au-Au
collisions at $\sqrt{s_{NN}}=200$~GeV - have been calculated
differentially in collision centrality within the PHSD transport
approach, which has been previously  tested in comparison to the
measured spectra and flow of photons in minimal bias collisions at
the same energy~\cite{Linnyk:2013hta}. We have found that the
channel decomposition of the photon spectra changes with centrality,
with a larger contribution of the hadronic sources in more
peripheral collisions.

As a consequence, the direct photon $v_2$ is larger in peripheral
collisions as compared to the most central reactions. We recall that
$v_2$ of photons produced in the initial hot deconfined phase by
partonic channels is small (of the order of 2\%) within our
approach~\cite{Linnyk:2013hta}. On the other hand, the photons from
the hadronic sources show strong elliptic flow (up to 10\%), on the
level of the $v_2$ of final hadrons~\cite{Linnyk:2013hta}.
Accordingly, since the channel decomposition of the direct photons
changes, their elliptic flow increases with decreasing centrality
and becomes roughly comparable with the $v_2$ of pions in peripheral
collisions. Moreover, the $v_2$ of the photons increases with
decreasing centrality additionally due to the rising of $v_2(b)$
with the impact parameter $b$, which was observed for all hadrons
(except for the most peripheral bin). The increase of the direct
photon $v_2(b)$ for the two most central bins has been also
indicated in hydrodynamics calculations in
Refs.~\cite{Shen:2013cca,Chaudhuri:2013eta}, although with slightly
lower absolute values of $v_2$. Future measurements of the photon
spectra and elliptic flow as a function of the collision centrality
will be mandatory for a clarification of the 'photon $v_2$ puzzle'
from the experimental side and to estimate the contribution from
unconventional sources
\cite{Bzdak:2012fr,Basar:2012bp,Goloviznin:2012dy,Yin:2013kya,Muller:2013ila,Skokov:2013axa}.

Additionally, our calculations have shown that the ``thermal" photon
$p_T$ spectra deviate from an exponential distribution at all
centralities primarily due to the hadronic bremsstrahlung channels.
The effective slopes of these spectra have been extracted in the
interval $p_T=(0.4-5)$~GeV and are constant with centrality within
error bars. Due to the non-exponential shape of the photon spectra
these effective slopes depend on the fitting interval in $p_T$,
however, provide 'effective temperatures' significantly above the
critical temperature $T_c \approx$ 160 MeV for the deconfinement
phase transition. Since in PHSD the dominant contributions to the
thermal photon yield are related to hadronic bremsstrahlung channels
the high slope parameters predominantly reflect the 'blue-shift' of
the photon spectra due to the collective flow of hadrons (cf. Ref.
\cite{Shen:2013vja}). Experimental data at low photon $p_T$ will
help in clarifying the physical sources.

Furthermore, since only collisional channels contribute to the
production of thermal photons in PHSD, their yield rises faster than
the number of participating nucleons $N_{part}$ as expected also
from qualitative considerations in
Refs.~\cite{Kajantie:1986dh,Srivastava:1999jz}. A power-law fit to
our results gives approximately a scaling $\sim N_{part}^\alpha$
with $\alpha \approx$1.5, whereas the partonic and hadronic channels
separately scale with exponents of $\approx$ 1.75 and $\approx$ 1.5,
respectively.

We finally point out that respective photon measurements of the
ALICE Collaboration at the LHC~\cite{Wilde:2012wc,Lohner:2012ct}
should complete the picture presented in this study. A detailed PHSD
analysis of photon production and flow at the LHC collision energies
will be reported in near future.

\section*{Acknowledgements}

The authors are grateful for fruitful discussions with B.~Bannier,
G.~David, C.~Gale, B.~Jacak, L.~McLerran, R.~Rapp, V.~Skokov,
A.~Toia, V.~Toneev and N.~Xu. This work was supported in part by the
LOEWE center HIC for FAIR.



\end{document}